\documentclass[
reprint,
 amsmath,amssymb,
 aps,
pra,
longbibliography,
]{revtex4-1}

\usepackage{graphicx}
\usepackage{dcolumn}
\usepackage{bm}
\usepackage{color}
\usepackage{hyperref}

\usepackage[english]{babel}
\makeatletter
\def\bbl@set@language#1{%
  \edef\languagename{%
    \ifnum\escapechar=\expandafter`\string#1\@empty
    \else\string#1\@empty\fi}%
  \@ifundefined{babel@language@alias@\languagename}{}{%
    \edef\languagename{\@nameuse{babel@language@alias@\languagename}}%
  }%
  \select@language{\languagename}%
  \expandafter\ifx\csname date\languagename\endcsname\relax\else
    \if@filesw
      \protected@write\@auxout{}{\string\select@language{\languagename}}%
      \bbl@for\bbl@tempa\BabelContentsFiles{%
        \addtocontents{\bbl@tempa}{\xstring\select@language{\languagename}}}%
      \bbl@usehooks{write}{}%
    \fi
  \fi}
\newcommand{\DeclareLanguageAlias}[2]{%
  \global\@namedef{babel@language@alias@#1}{#2}%
}
\makeatother
\DeclareLanguageAlias{en}{english}

\begin{document}

\title{Spin and Valley States in Gate-defined Bilayer Graphene Quantum Dots}

\author{Marius Eich}
 \email{meich@phys.ethz.ch}

\author{Riccardo Pisoni}
\author{Hiske Overweg}
\author{Annika Kurzmann}
\author{Yongjin Lee}
\author{Peter Rickhaus}
\author{Thomas Ihn}
\author{Klaus Ensslin}%
 
\affiliation{%
 Solid State Physics Laboratory, ETH Zurich, 8093 Zurich, Switzerland \\
}

\author{Franti\v{s}ek Herman}
\author{Manfred Sigrist}
\affiliation{
 Institute for Theoretical Physics, ETH Zurich, 8093 Zurich, Switzerland \\
}

\author{Kenji Watanabe}
\author{Takashi Taniguchi}
\affiliation{%
 Advanced Materials Laboratory, NIMS, 1-1 Namiki, Tsukuba 305-0044, Japan \\
}

\date{\today}

\begin{abstract}
In bilayer graphene, electrostatic confinement can be realized by a suitable design of top and back gate electrodes. 
We measure electronic transport through a bilayer graphene quantum dot, which is laterally confined by gapped regions and connected to the leads via p-n junctions. 
Single electron and hole occupancy is realized and charge carriers $n = 1, 2,\dots 50$ can be filled successively into the quantum system with charging energies exceeding $10 \ \mathrm{meV}$. 
For the lowest quantum states, we can clearly observe valley and Zeeman splittings with a spin g-factor of $g_{s}\approx 2$. 
In the low field-limit, the valley splitting depends linearly on the perpendicular magnetic field and is in qualitative agreement with calculations.
\end{abstract}

\maketitle

\section{\label{sec:Intro}Introduction}

Graphene has been recognized early on as a prime candidate to host spin qubits \cite{trauzettel_spin_2007}. 
With carbon being one of the lightest elements in the periodic table, spin-orbit interactions are expected to be weak. In addition, 99\% of natural carbon consists of nuclear spin-free $^{12}$C. 
Therefore, the two main spin decoherence mechanisms for spin qubits, namely spin-orbit interactions and hyperfine coupling of nuclear and electronic spins, should be strongly suppressed in any carbon-based solid state system. 
So far, these theoretical considerations have not come to fruition in experiments.

Until now, graphene quantum dots (QDs) have been mostly realized by top-down lithography and etching of single layer graphene \cite{stampfer_tunable_2008-1,ponomarenko_chaotic_2008,liu_gate-defined_2010,guttinger_transport_2012,bischoff_localized_2015}. 
While many of the basic quantum transport properties such as Coulomb blockade  \cite{stampfer_tunable_2008-1,ponomarenko_chaotic_2008}, charge detection \cite{guttinger_charge_2008}, and electronic phase coherence \cite{miao_phase-coherent_2007,russo_observation_2008} have been experimentally demonstrated, the understanding of the orbital and spin character of specific states has remained elusive. 
In retrospect, we understand that Coulomb blockade in these devices arises mostly from localized states at the sample edges, which remain rough on the atomic scale because of limitations of top-down technology\cite{bischoff_localized_2015}. 

More than a decade ago, experiments have shown that a band gap can be opened in bilayer graphene by vertical electric fields \cite{mccann_asymmetry_2006,ohta_controlling_2006,oostinga_gate-induced_2008} and charge carrier confinement in bilayer graphene has been studied in theory \cite{pereira_tunable_2007,recher_bound_2009,zarenia_electron-electron_2013}.
Several attempts to use split-gate electrodes to laterally confine charge carriers in the absence of a magnetic field have suffered from limited resistance values that can be experimentally obtained upon pinch-off and did not reach the last-electron regime \cite{allen_gate-defined_2012,goossens_gate-defined_2012,zhu_edge_2017}. 
Recently, we realized quantum point contacts that display quantized conductance and show pinch-off resistances orders of magnitude larger than the quantum of resistance $h/e^{2}$ \cite{overweg_electrostatically_2018}, a necessary requirement to electrically isolate charge carriers from their environment. 
Here, the same fabrication technique has been adapted using suitable gate geometries to prepare QDs with an electronic quality that matches what has been achieved in the traditional semiconductors Si and GaAs \cite{zwanenburg_silicon_2013,kouwenhoven_few-electron_2001}.

\begin{center}
\begin{figure*}[htb]
\includegraphics[width=0.9\textwidth]{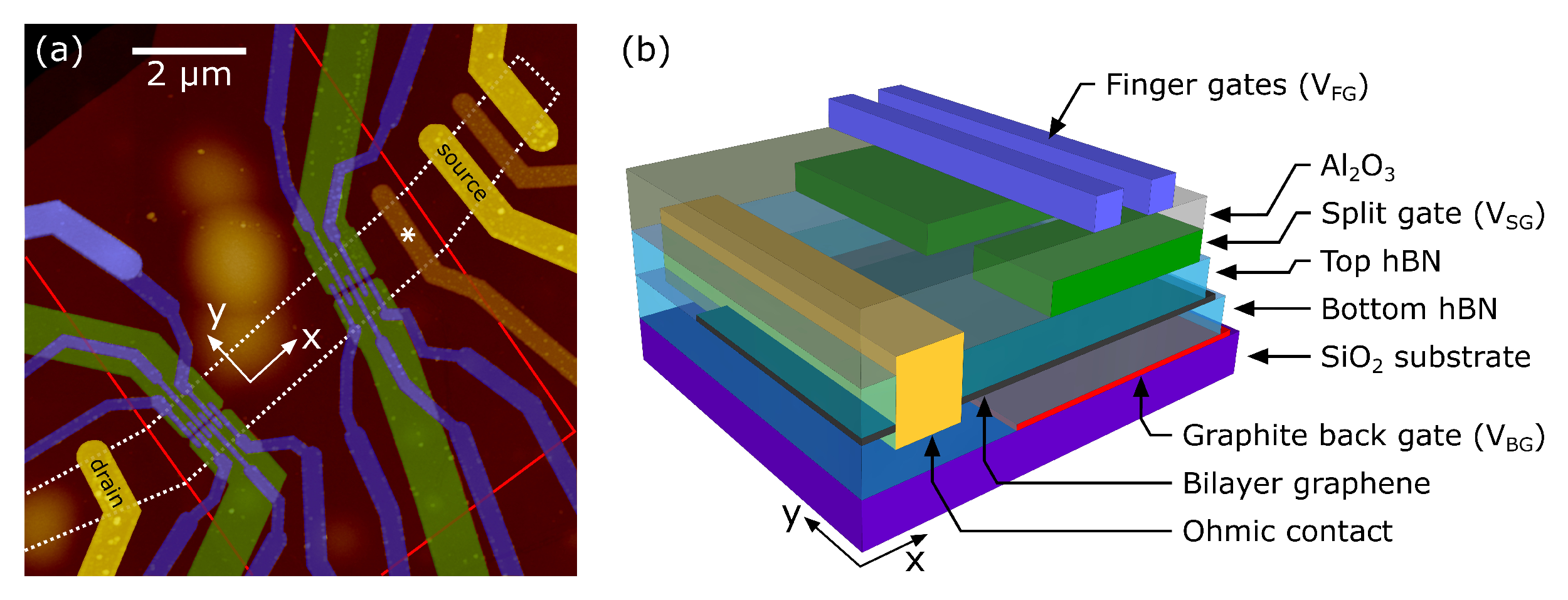}
\caption{
(a)~False color scanning force micrograph of the device. 
(b)~3D sketch of part of the device showing the different layers of gates and dielectrics. White dashed lines and solid red lines in (a) outline the bilayer flake and graphite back gate, respectively. Edge contacts to the bilayer are colored in yellow, split gates are shown in green and the finger gates are shown in blue. The top gate spanning the entire width of the bilayer flake is marked with an asterisk.
}
\end{figure*}
\end{center}

\vspace{-7ex}
In the first experiment we demonstrate charging of a bilayer graphene QD with a single and a few holes when coupling the QD to n-type source and drain leads through p-n tunnel barriers. 
Charging energies in excess of 10 meV are observed. 
We reverse the gate voltages and investigate single or few electron QDs connected to p-type leads, demonstrating the ambipolar operation of these QDs on the same graphene flake in close vicinity to each other.
Applying perpendicular magnetic fields in the second experiment, we extract the single particle level spectrum, showing shell filling and orbital degeneracy. 
The pronounced valley splitting is in agreement with calculations, which predict that the splitting depends on the dot size. 
In the third experiment, we carefully align the graphene sheet hosting the QD to an in-plane magnetic field and find a Zeeman splitting with a g-factor $g_{s} = 2.08 \pm 0.22$ agreeing with the expected value for carbon-based devices \cite{thess_crystalline_1996,tans_individual_1997}.

Our bilayer QDs display high-quality electronic properties comparable to standard semiconductor structures that have been optimized for the last 30 years. 
While excellent devices have also been reported for carbon nanotube QDs \cite{cobden_shell_2002,jarillo-herrero_electron-hole_2004,sapmaz_quantum_2006,steele_large_2013,laird_valleyspin_2013}, graphene offers the advantage of a planar technology \cite{kouwenhoven_few-electron_2001} and the possible combination with other 2D materials \cite{geim_van_2013,novoselov_2d_2016,mcdonnell_atomically-thin_2016}. 
Our demonstration of excellent control and reproducibility opens up a wide field of possibilities for carbon-based quantum electronics.

\section{\label{sec:Char}Characterization}

We investigated the bilayer graphene device encapsulated in hexagonal boron nitride \cite{dean_boron_2010,wang_one-dimensional_2013,uwanno_fully_2015} shown in Fig.~1. 
The individual layers of the van der Waals heterostructure were stacked and processed as in ref.~\cite{overweg_electrostatically_2018}, protecting the natural edges of the bilayer flake (white dashed lines in Fig.~1(a)). 
Opposite voltages applied to the split gates (green in Fig.~1) and the graphite back gate (red solid lines in Fig.~1(a)) lead to the formation of a band gap in the bilayer regions underneath the split gates. 
For appropriate voltages applied to theses gates, the Fermi level is tuned to be in the band gap (for details, see ref.~\cite{overweg_electrostatically_2018}), rendering these regions insulating and defining $\sim 100 \ \mathrm{nm}$ wide channels between the source and drain contacts (contacts shown in yellow in Fig.~1).
Finger gates (blue in Fig.~1, numbered 1 through 11 in x-direction, see Fig.~1(a)) crossing the channel on top of the two split gate pairs (insulated from them by $25 \ \mathrm{nm}$ of Al$_{2}$O$_{3}$) are biased to control the charge carrier density locally in the channel.

First, we investigate the conductance of the device biasing only the uniform top gate crossing the entire width of the bilayer region (white asterisk in Fig.~1(a)). 
By applying large opposite voltages to the graphite back gate and this top gate, the strong displacement field opens a band gap in the bilayer region underneath the top gate. 
The two-terminal resistance measured between the source and drain contacts reaches values on the order of $\mathrm{G\Omega}$ when tuning the Fermi level into the gap (see \href{https://journals.aps.org/prx/abstract/10.1103/PhysRevX.8.031023}{Supplementary Material}), demonstrating the high electronic quality of our sample and the excellent insulating behavior of the gapped region. 
Biasing either pair of split gates in this regime of high displacement field, charge carriers are laterally confined and forced to flow through the narrow channel between the split gates. 
This is the regime in which we form and operate our QDs.
\vspace{1ex}

\section{\label{sec:Res}Results}

\subsection{\label{subsec:QD}Gate-defined quantum dots}

\begin{center}
\begin{figure*}[htb]
\includegraphics[width=0.9\textwidth]{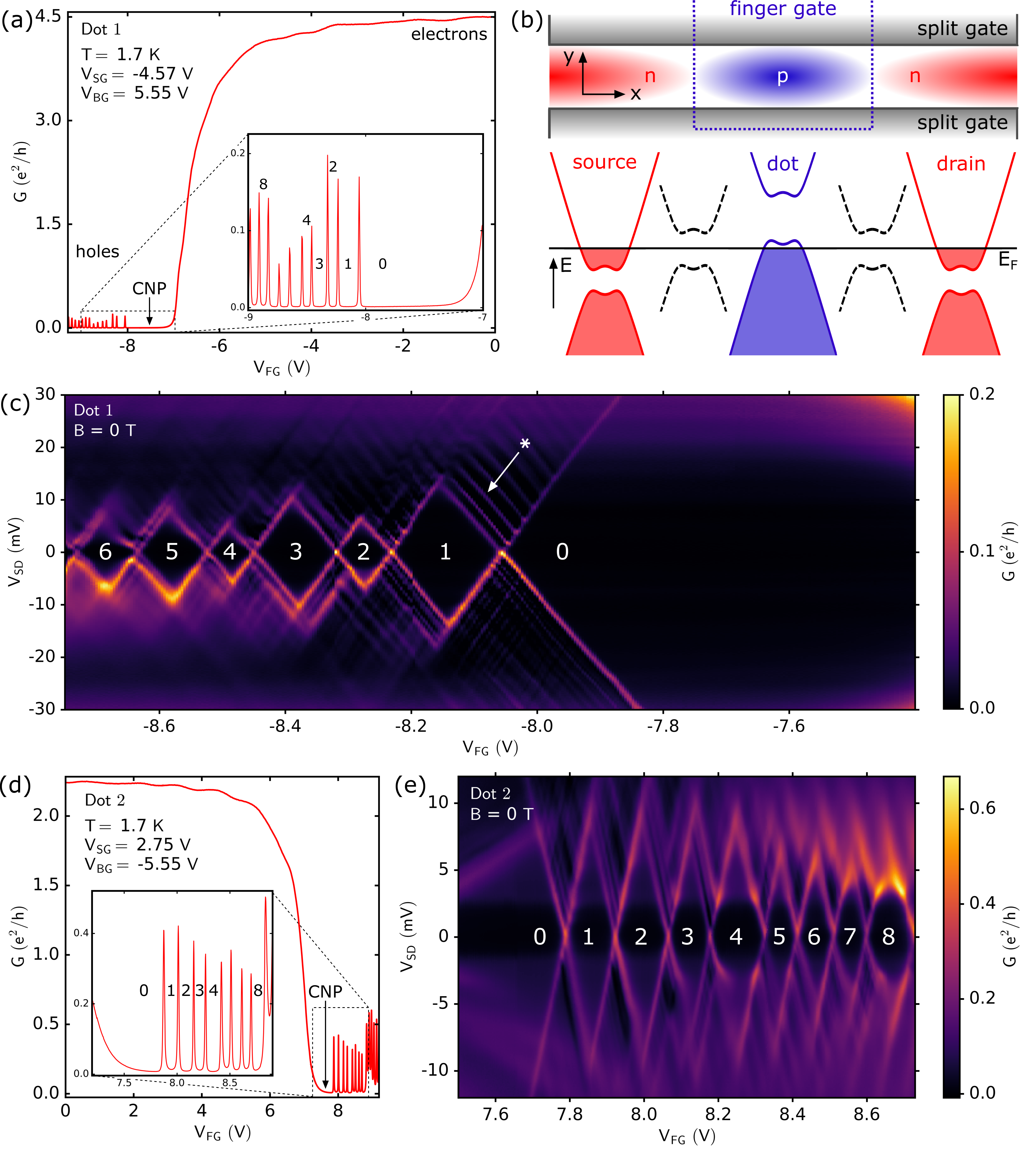}
\caption{
(a)~Conductance trace for p-type QD 1. 
(b)~Schematic of the band structure at different positions along the current direction in the channel. Black dashed lines schematically show the band alignment between the dot region and the leads.
(c)~Coulomb diamonds in the hole regime for QD 1 with an asterisk indicating regularly spaced resonances parallel to the diamond boundaries.
(d)~Conductance trace and (e) Coulomb diamonds for the n-type QD 2. Numbers in (a) and (c), as well as (d) and (e) indicate the occupation of the QDs with holes or electrons, respectively.
}
\end{figure*}
\end{center}

\vspace{-4ex}
In the first experiment at $1.7 \ \mathrm{K}$ we investigate charging a QD with single holes. 
We measured nine QDs in total, all showing qualitatively the same behavior. 
By recording conductance maps as function of finger gate and split gate voltage, a particular QD can be tuned to an optimal operation point (see \href{https://journals.aps.org/prx/abstract/10.1103/PhysRevX.8.031023}{Supplementary Material}). 
Figure~2(a) shows the conductance of the device as a function of the finger gate voltage $V_{FG}$. 
Charge carriers can only flow through the narrow channel, because the regions underneath the split gates are insulating. 
The positive back gate voltage $V_{BG}$ induces a finite excess electron density in the channel. 
With decreasing finger gate voltage $V_{FG}$ the electron density is locally reduced until complete pinch-off is reached at the charge neutrality point (CNP) at about -7.5 V.

\begin{center}
\begin{figure*}[htb]
\includegraphics[width=0.9\textwidth]{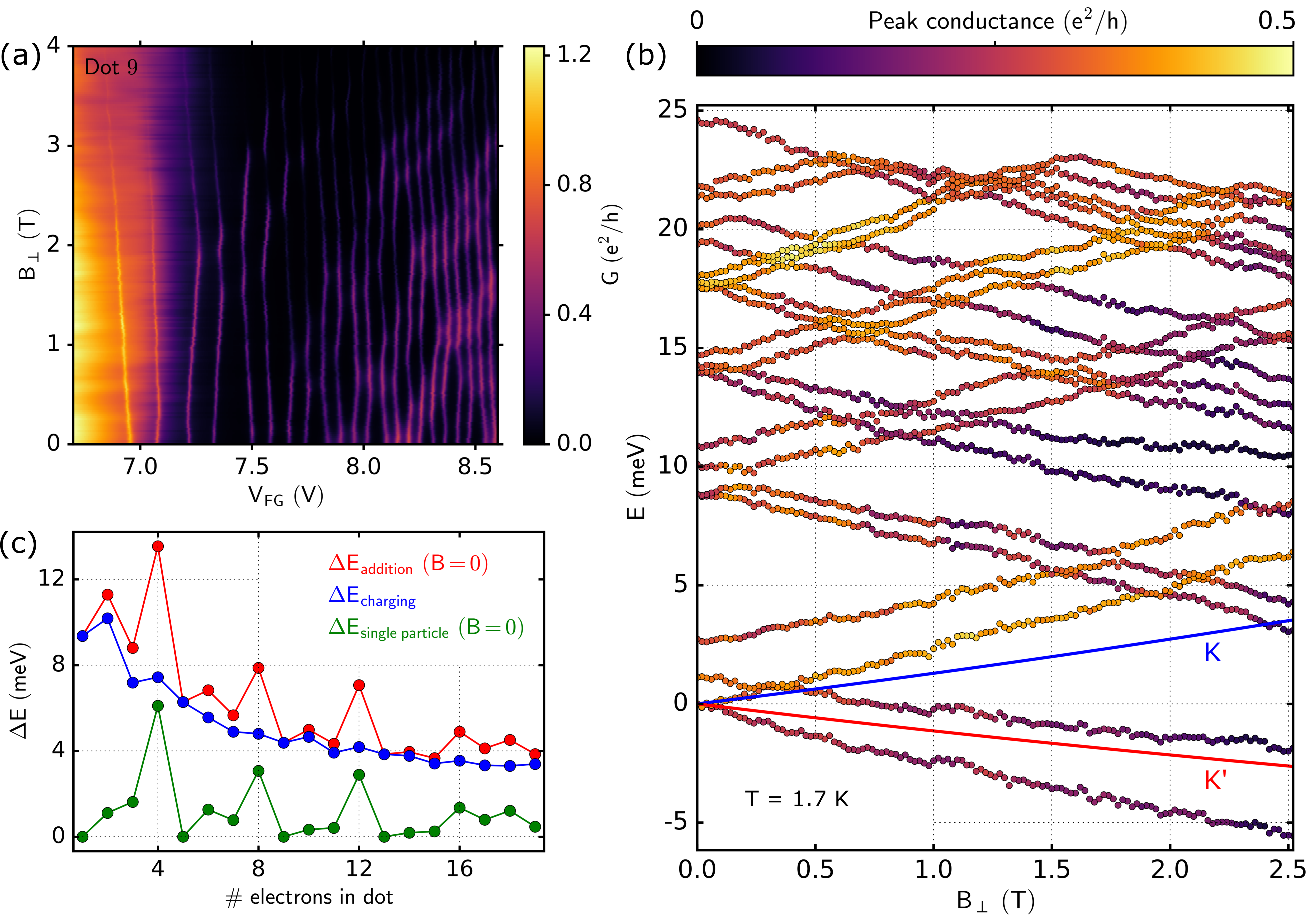}
\caption{
(a)~Conductance map in a perpendicular magnetic field for QD 9 in the electron regime. 
(b)~Single particle energy level dispersion for QD 9 with perpendicular magnetic field $B_{\perp}$ extracted from Fig.~3(a). Blue and red solid lines are the result of theoretical calculations of the lowest energy levels. 
(c)~Addition (red), charging (blue) and single particle energy (green) as function of the QD occupation extracted from Fig.~3(a).
}
\end{figure*}
\end{center}

\vspace{-4.5ex}
For $V_{FG} < -7.5 \ \mathrm{V}$ the region underneath the finger gate is tuned into the hole regime as shown schematically in Fig.~2(b): 
We sketch at the top the n-type channel (red) with the locally induced p-type region (blue). Below we show the dispersion relation near the K-point of the first Brillouin zone in the three spatial regions and at the p-n junctions between them (dashed).
At the p-n junctions, the Fermi level $E_{F}$ lies in the gap leading to a region with zero charge carrier density. 
These regions provide natural tunnel barriers separating the p-type dot from the n-type leads. 
By lowering the finger gate voltage in this regime, the p-type QD can be charged one-by-one with individual holes. 
This is seen in Fig.~2(a) at $V_{FG} < -7.5 \ \mathrm{V}$, where sharp conductance resonances appear (see also inset).

Finite DC bias spectroscopy of the QD tuned to this regime yields the Coulomb diamonds shown in Fig.~2(c). 
For $V_{FG} > -8 \ \mathrm{V}$ we do not see additional states contributing to transport through the QD, indicating a completely depleted dot. 
Therefore we label each diamond with the occupation number of the QD (c.f. inset of Fig.~2(a)). 
The regularly spaced features running parallel to the edge of the Coulomb diamonds (indicated by an asterisk in Fig.~2(c)) appear for all measured QDs, are stable over a temperature range from $50 \ \mathrm{mK}$ to $1.7 \ \mathrm{K}$ and are currently still under investigation.

To form an electron QD connected to p-type leads, we reverse all applied gate voltages with respect to the overall charge neutrality point. 
The conductance trace in Fig.~2(d) mirrors the situation of the hole QD in Fig.~2(a). 
The n-type QD can also be charged one-by-one with individual electrons, proving the ambipolar operation of our bilayer QDs. 
The corresponding Coulomb diamonds for the electron dot (Fig.~2(e)) again mirror the situation of Fig.~2(c). 
In the electron as well as in the hole regime charging energies on the order of $10 \ \mathrm{meV}$ are observed. 
In contrast to the p-type QD presented in Fig.~2(c), the n-type QD exhibits additional features in the region of zero charge carrier occupation of the QD.
These features depend on the precise setting of the split gate voltage $V_{SG}$ and correspond to localized states in the leads close to the QD (see Fig.~S2).
In total, 8 different QDs were measured both in the electron and hole regime, all showing qualitatively the same results.

\subsection{\label{subsec:Lev}Level structure}

In the second experiment at $1.7 \ \mathrm{K}$, we measure Coulomb resonances of the electron QD 9 as a function of a perpendicular magnetic field $B_{\perp}$ (conductance map in Fig.~3(a)). 
The shifts of the resonances in $V_{FG}$ as a function of $B_{\perp}$ correspond to shifts of energy levels of the QD evolving with $B_{\perp}$.
To extract the energy level spectrum of our QD from the resonance spacings, we subtract the charging energy \cite{fuhrer_energy_2001,kouwenhoven_few-electron_2001,ihn_semiconductor_2009} by shifting neighboring resonances such that they touch in a single point and convert the voltage to an energy axis (see \href{https://journals.aps.org/prx/abstract/10.1103/PhysRevX.8.031023}{Supplementary Material}).
The extracted magnetic field-dependence of the energy levels is shown in Fig.~3(b), where the color scale indicates the peak conductance of each level which is proportional to the coupling of the corresponding state to the leads \cite{ihn_semiconductor_2009}.

We see that the levels bunch in groups of four at zero magnetic field, as expected from the two-fold valley and two-fold spin degeneracy in bilayer graphene, and similar to carbon nanotubes \cite{cobden_shell_2002,sapmaz_quantum_2006}. 
Each shell of four states splits into 2+2 as the magnetic field is increased, one pair shifting up, the other down in energy. 
The splitting between these states is linear in the accessible magnetic field range, and is forty times stronger than the Zeeman splitting for a free electron.

To compare the data with theory we used the bilayer QD model presented in \cite{recher_bound_2009} and adapted it to our system. 
The allowed energy levels depend on the valley index (labeled by $\tau = \pm 1$ in the theory corresponding to the K and K' valley), the angular momentum number $m$, the interlayer asymmetry $V$, the confinement potential $U$, and the radius $R$ of the QD. 
The levels have to be calculated numerically by matching four-component spinor states at the QD boundary. 
From the displacement field applied in the experiment, the interlayer asymmetry was estimated to be $V = 60 \ \mathrm{meV}$ \cite{mccann_electronic_2013}. 
Since electrons are confined electrostatically, the confinement potential $U$ should be on the order of the interlayer asymmetry $V$ and we fix $U=V$. 
The remaining parameter $R$ determines both the orbital energy level difference at $B_{\perp} = 0$ as well as the valley splitting as function of $B_{\perp}$. 
To reproduce the observed orbital energy level difference on the order of $5 \ \mathrm{meV}$ we obtain $R = 20 \ \mathrm{nm}$ which agrees well with the lithographic design of the device. 
The calculated energy levels (spin-degenreate in the theory) are shown in blue and red in Fig.~3(b) for the K and K' states, respectively. 
To improve the theoretical model it should be adapted to the non-radial symmetry of the dot and the non-homogeneous confinement laterally and in transport direction (see \href{https://journals.aps.org/prx/abstract/10.1103/PhysRevX.8.031023}{Supplementary Material}).
The experimentally observed valley splitting varies by $20\%$ between different QDs which could be a result of the microscopic differences in the size of the individual QDs caused in fabrication.

\begin{center}
\begin{figure}[htb]
\includegraphics[width=0.45\textwidth]{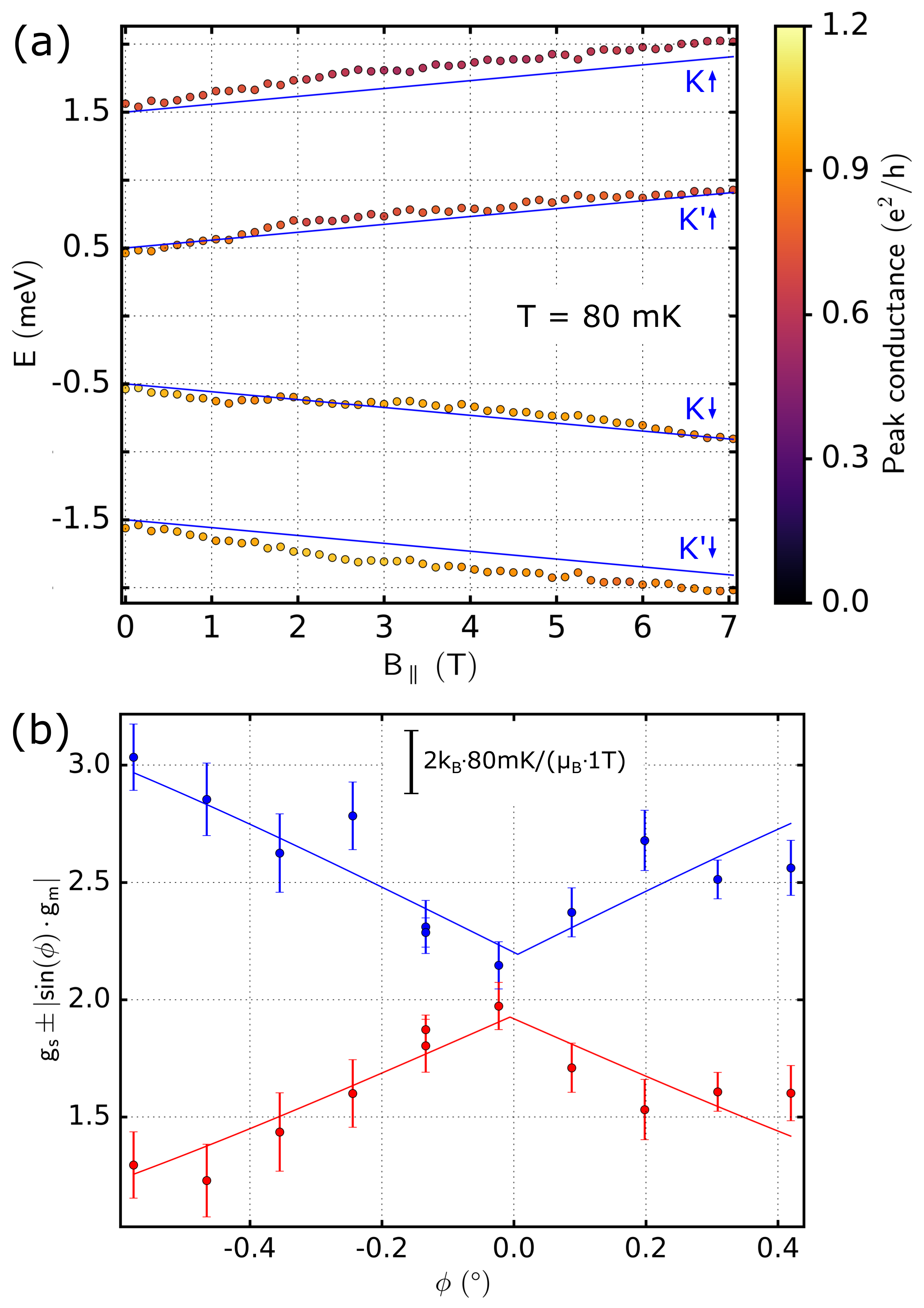}
\caption{
(a)~Lowest single particle energy levels of QD 2 in the electron regime as a function of parallel magnetic field. Energy levels are shifted by $1 \ \mathrm{meV}$ for clarity. 
(b)~Sum of the spin Zeeman ($g_{s}$) and orbital ($g_{m}$) contribution to the energy level splitting as function of the sample tilt angle $\phi$ with respect to magnetic field. Blue (red) points are averaged over levels K'$\downarrow$ and K$\uparrow$ (levels K$\downarrow$ and K'$\uparrow$) and solid lines represent fits to the data.
}
\end{figure}
\end{center}

\vspace{-5ex}
Figure~3(c) shows the experimental spacing of neighboring states as a function of occupation number. 
The single particle level spacing at zero field and the charging energy are shown in green and blue, respectively. 
The charging energy decreases with increasing number of electrons occupying the QD because the effective electronic dot gets larger \cite{kouwenhoven_few-electron_2001,ihn_semiconductor_2009}. 
The addition energy, the sum of the two former, is directly proportional to the spacing of Coulomb resonances in finger gate voltage at zero field. 
We observe a clear four-fold level bunching (Figs.~3(b),(c)) originating from the two-fold spin and valley degeneracy of bilayer graphene \cite{pereira_tunable_2007,recher_bound_2009,zarenia_electron-electron_2013}, which can already be seen in Figs.~2(d),(e). 
Until now, this intrinsic property specific to graphene QDs has not been observed experimentally. 
The same level bunching was also observed for the hole QDs at $35 \ \mathrm{mK}$ (see \href{https://journals.aps.org/prx/abstract/10.1103/PhysRevX.8.031023}{Supplementary Material}).

\subsection{\label{subsec:gfac}g-factor}

In the third experiment at $80 \ \mathrm{mK}$, we align the device parallel to the magnetic field in a revolving sample holder. 
The energy levels in Fig.~4(a) as a function of the parallel magnetic field are extracted in the same way as for Fig.~3(b), but vertically shifted by $1 \ \mathrm{meV}$ for clarity. 
Blue lines are guides to the eye for purely Zeeman split energy levels with a spin g-factor of $g_{s} = 2$. 
Repeating the measurement of Fig.~4(a) for different rotation angles $\phi$ enables us to extract the spin Zeeman and orbital contribution to the splitting of energy levels as a function of the angle. 
The orbital contribution adds to the Zeeman splitting and is proportional to the perpendicular component of the magnetic field. 
With respect to horizontal lines in Fig.~4(a), the splitting is enhanced for levels K'$\downarrow$ and K$\uparrow$ and is reduced for levels K$\downarrow$ and K'$\uparrow$. 
Averaging over these pairs of levels leads to the data shown in Fig.~4(b) where solid lines represent fits to the data. 

In theory, the red and blue points should touch at $g_{s} = 2$ for perfect alignment of the sample parallel to the magnetic field and the slope of the fits should correspond to the slopes of the lowest four levels Fig.~3(b). 
The extracted orbital splitting for $\phi = 90 ^{\circ}$ is 35 (blue) to 40 (red) times stronger than the Zeeman effect for a free electron, matching the experimental data from Fig.~3(b).
The extracted spin g-Factor of $g_{s} = 2.08 \pm 0.22$ also agrees well with the predicted value for carbon-based devices \cite{thess_crystalline_1996,tans_individual_1997}.

With the experiments in parallel and perpendicular magnetic fields we can show that the observed four-fold level bunching originates from the two-fold spin degeneracy (split in $B_{\parallel}$) and the two-fold valley degeneracy (split in $B_{\perp}$) of bilayer graphene.

Follow-up manuscripts reporting on double- and mulit-QD systems in bilayer graphene have meanwhile appeared in the literature \cite{eich_coupled_2018,banszerus_gate-defined_2018}.

\section{\label{sec:Conc}Conclusion}

The presented experimental results and the qualitative agreement with theoretical calculations prove the quality and understanding of our bilayer QDs. 
The electrostatic confinement of single charge carriers in a planar technology is an important step toward the promising implementation of spin and valley qubits in graphene-based devices. 
With the p-n junctions serving as natural tunnel barriers, QDs can be coupled in the future to create a series of QDs with alternating polarity. 
We expect that the implementation of high-frequency read-out will enable the determination of spin and valley coherence in graphene quantum dots and open up new horizons for spin and valley qubit research.

\acknowledgments

We thank Jelena Klinovaja and Francois Peeters for fruitful discussions, and Peter M\"arki, Erwin Studer, as well as the FIRST staff for their technical support.
We also acknowledge financial support from the European Graphene Flagship, the Swiss National Science Foundation via NCCR Quantum Science and Technology, the EU Spin-Nano RTN network and ETH Zurich via the ETH fellowship program. Growth of hexagonal boron nitride crystals was supported by the Elemental Strategy Initiative conducted by the MEXT, Japan and JSPS KAKENHI Grant Number JP15K21722.

\noindent \textbf{Authors contributions:} M.E. and R.P. fabricated the device. M.E. performed the measurements. F.H. performed the theoretical calculations. H.O., A.K., Y.L., and P.R. supported device fabrication and data analysis. K.W. and T.T. provided high-quality boron nitride crystals. K.E., T.I., and M.S. supervised the work. The authors declare that they have no competing interests.

\noindent \textbf{Supplementary Material:} The Supplementary Material is included in the published PRX manuscript and is available free of charge on the \href{https://journals.aps.org/prx/abstract/10.1103/PhysRevX.8.031023}{PRX website} including sections on materials and methods, supporting data and theoretical considerations.

\end{document}